\documentclass[prb,twocolumn,amsmath,amssymb,superscriptaddress,floatfix,nofootinbib]{revtex4-2}

\usepackage{graphicx}
\usepackage{dcolumn}
\usepackage{bm}
\usepackage{ textcomp }
\usepackage{color}
\usepackage{amsmath}
\usepackage{amssymb}
\usepackage{xcolor}
\usepackage[colorlinks = true,
            linkcolor = blue,
            urlcolor  = red,
            citecolor = blue,
            anchorcolor = blue]{hyperref}
\usepackage{easyReview}
\usepackage{mathdots}
\usepackage{float}
\usepackage{lineno}
\usepackage{todonotes}
\usepackage{url}
\usepackage{array}
\usepackage{todonotes}
\usepackage{xcolor,colortbl}

\usepackage{lipsum, babel}

\definecolor{LightBlue}{rgb}{0.8,0.8,0.8}

\usepackage[normalem]{ulem}

\begin{document}

\title{Weak-Coupling Theory of Neutron Scattering as a Probe of Altermagnetism}

\date{\today}

\author{Thomas A. Maier} \affiliation{Computational Sciences and Engineering Division, Oak Ridge National Laboratory, Oak Ridge, Tennessee 37831, USA}

\author{Satoshi Okamoto} \affiliation{Materials Science and Technology Division, Oak Ridge National Laboratory, Oak Ridge, Tennessee 37831, USA}

\begin{abstract}
Inelastic neutron scattering provides a powerful probe of the magnetic excitations of quantum magnets. Altermagnets have recently emerged as a new class of magnets with vanishing net magnetization characteristic of antiferromagnets and with a spin-split electronic structure typical of ferromagnets. Here we introduce a minimal Hubbard model with two-sublattice orthorhombic anisotropy as a framework to study altermagnetism. Using unrestricted Hartree-Fock calculations, we find an altermagnetic state for this model that evolves from a metallic state to an insulating state with increasing Hubbard-$U$ Coulomb repulsion. We then examine the inelastic neutron scattering response in these states using random-phase approximation calculations of the dynamic spin susceptibility $\chi''({\bf q}, \omega)$. We find that the magnetic excitation spectrum depends on its chirality for ${\bf q}$ along certain directions in reciprocal space, an observation that may be used in inelastic neutron scattering experiments as a probe of altermagnetism. 
\end{abstract}

\maketitle

\section{Introduction} 

Altermagnetism has emerged as a new class of magnetism \cite{Smejkal2022a,Smejkal2022b,Hayami2019,Hayami2020}. As conventional antiferromagnets, altermagnets have an alternating spin polarization in real space. In contrast to antiferromagnets, however, the magnetic ordering in altermagnets breaks the $\cal PT$ (parity times time-reversal) symmetry. As a consequence, the spin splitting also appears in reciprocal space in the electronic band structure. A large number of candidate materials have already been identified to host altermagnetism, including RuO$_2$ \cite{Berlijn2017,Ahn2019,Smejkal2020}, FeSb$_2$ \cite{Mazin2022}, and other materials \cite{Smejkal2022a,Smejkal2022b}. When rotational ${\cal C}_4$ symmetry is preserved, such as in RuO$_2$, the band-dependent spin polarization has $d_{x^2-y^2}$-symmetry, and the analogy with unconventional $d$-wave superconductivity as observed in the cuprates has been pointed out \cite{Smejkal2022b}. Moreover, based on their unique electronic structure, several phenomena have been proposed that are especially useful for spintronics applications with itinerant electrons \cite{Smejkal2022a,Smejkal2022b,Bose2022,Feng2022}. 
 
While the unique electronic structure of altermagnets has received most attention, especially in the context of spintronics applications, the momentum dependent spin splitting and spin dynamics have just now started to become the focus of research. For example, Ghorashi {\it et al.} proposed a heterostructure consisting of an altermagnet and a superconductor as a platform to create Majorana zero modes with net zero magnetization \cite{Ghorashi2023}. Steward {\it et al.} used a phenomenological model for the spin dynamics to propose a way to assess the spin excitations in altermagnets directly from the phonon spectrum utilizing the coupling between spin excitations and lattices degree of freedom \cite{Steward2023}. Moreover, {\v S}mejkal {\it et al.} theoretically examined the spin excitations in RuO$_2$ \cite{Smejkal2022c} based on strong coupling linear spin-wave theory, and found an alternating chirality splitting in the magnon bands different from the chirality-degenerate magnon bands in conventional antiferromagnets. Many candidate altermagnets, however, are weakly correlated itinerant systems, such as RuO$_2$ for example, and especially the materials that are more favorable for practical applications in spintronics. In that case, a more natural starting point is a weak coupling theory based on an itinerant picture. 

Here we use weak coupling theory to examine the spin dynamics in altermagnets and explore how inelastic neutron scattering in the magnetically ordered state can be used to distinguish between a conventional antiferromagntic and an altermagnetic state. To this end, we introduce a minimal microscopic Hubbard model with an on-site Coulomb $U$ repulsion for an altermagnet, and a corresponding model for an (anisotropic) antiferromagnet. We first study the phase diagram of these models for varying $U$ using Hartree Fock (HF) theory. Next, we proceed to explore the structure of the inelastic neutron scattering in the magnetic state for both altermagnetic and antiferromagnetic models using a weak-coupling random phase approximation. 

\section{Model} 

\begin{figure}[t!] 
  \includegraphics[width=0.45\textwidth]{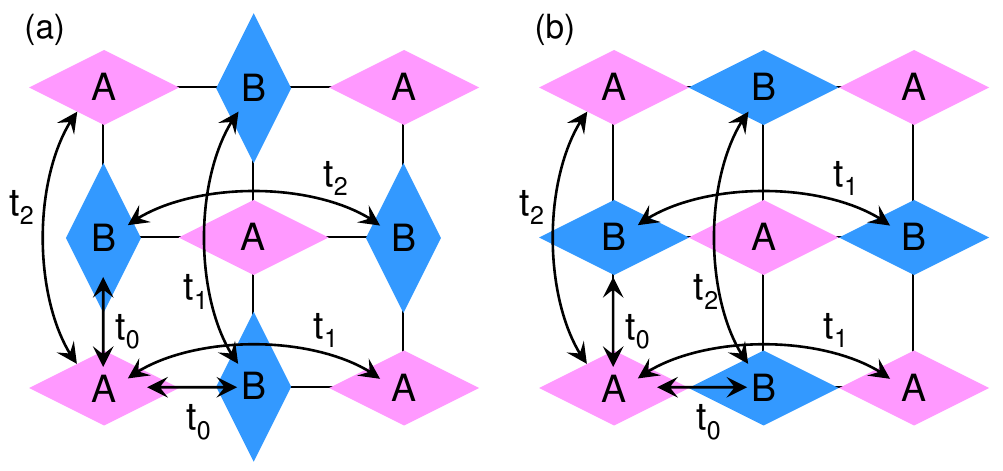} 
  \caption{Illustration of the models considered in the calculations: (a) altermagnet when sublattices A and B have opposite spin polarizations and (b) antiferromagnet which has an orthorhombic anisotropy. \label{fig:1}} 
\end{figure}

To model the altermagnetic state, we start with a two-dimensional (2D) Hubbard model on a square lattice with the total number lattice sites $L$. Its Hamiltonian
\begin{eqnarray} \label{eq:H}
	H = -\sum_{ij, \sigma}(t_{ij}+\mu\delta_{ij}) c^\dagger_{i\sigma}c^{\phantom\dagger}_{j\sigma} + U\sum_i n_{i\uparrow} n_{i\downarrow}\,,
\end{eqnarray}
is formulated in terms of the usual electron creation ($c^\dagger_{i\sigma}$) and annihilation ($c_{i\sigma}$) operators for sites $i$ and spin $\sigma$, their corresponding number operators $n_{i\sigma}=c^\dagger_{i\sigma}c^{\phantom\dagger}_{i\sigma}$, the single-particle hopping amplitude $t_{ij}$ between sites $i$ and $j$, 
and the on-site Coulomb repulsion $U$. $\mu$ is the chemical potential. As shown in Fig.~\ref{fig:1}, panels (a) and (b), the 2D square lattice is divided into A- and B-sublattices to allow for different intra-sublattice (A-A and B-B) hopping terms. We include three terms for the hopping amplitude $t_{ij}$: A nearest neighbor (inter-sublattice A-B) term $t_0$ and third-nearest-neighbor (intra-sublattice A-A and B-B) terms $t_1$ and $t_2$, which break the sublattice ${\cal C}_4$ symmetry. The minimal model we propose to describe the altermagnetic state is illustrated in Fig.~\ref{fig:1} (a). On sublattice A, it has intra-sublattice hopping $t_1$ along $x$ and $t_2$ along $y$. On the B-sublattice, $t_1$ and $t_2$ are switched. Fourier-transforming the first (non-interacting) term of the Hamiltonian in Eq.~(\ref{eq:H}) to  reciprocal space gives 
\begin{equation}
	H_0 = 
	\sum_{{\bf k},\sigma} 
	\psi^\dagger_{{\bf k}\sigma} {\bf h}({\bf k}) \psi^{\phantom\dagger}_{{\bf k}\sigma}\,.
\end{equation}
Here, 
${\bf k}$ is the crystal momenta of a sublattice, $\psi^\dagger_{{\bf k}\sigma} = (c^\dagger_{A{\bf k}\sigma}, c^\dagger_{B{\bf k}\sigma})$ and 
\begin{eqnarray} \label{eq:hk}
  {\bf h}({\bf k}) &=& \frac{1}{2}[\epsilon_{AA}({\bf k})+\epsilon_{BB}({\bf k})]\tau_0 + \epsilon_{AB}({\bf k})\tau_1 \nonumber\\
   &+& \frac{1}{2}[\epsilon_{AA}({\bf k})-\epsilon_{BB}({\bf k})]\tau_3 - \mu\tau_0
\end{eqnarray}

with
\begin{eqnarray}
	\epsilon_{AA}({\bf k}) &=& -2t_1\cos 2k_x-2t_2\cos 2k_y , \\
	\epsilon_{BB}({\bf k}) &=& -2t_2\cos 2k_x-2t_1\cos 2k_y , \\
	\epsilon_{AB}({\bf k}) &=& -2t_0 (\cos k_x+\cos k_y) \,.
\end{eqnarray}
and $\tau_i$ the Pauli matrices in sublattice space. 
As we will see, the bandstructure of this model leads to an altermagnetic state  that breaks $\cal PT$ symmetry. 

For comparison, we will also consider an analogous model with translationally invariant hopping amplitudes that supports a conventional antiferromagnetic state. As illustrated in Fig.~\ref{fig:1} (b), this model has the same $t_1$ and $t_2$ hopping terms on both A- and B-sublattices so that $\epsilon_{BB}({\bf k}) = \epsilon_{AA}({\bf k})$. In this case, we will see that its magnetic bandstructure is two-fold (spin) degenerate and therefore does not violate the $\cal PT$ symmetry. 

In the following we take $t_0=1$ as the unit of energy and set $t_1=0.4t_0$ and $t_2=0.2t_0$. The chemical potential $\mu$ is adjusted so that the model is half-filled with an average site filling $\langle n\rangle$ = 1.

\section{Hartree-Fock phase diagram} 

We start by analyzing the models shown in Fig.~\ref{fig:1} with Hamiltonian in Eq.~(\ref{eq:H}) 
using an unrestricted Hartree-Fock approximation. 
This approximation allows one to investigate possible symmetry-broken phases of interacting models 
and excitations from such phases. 

To proceed, we introduce a mean-field decoupling in the interaction term of Eq.~(\ref{eq:H}) as 
\begin{eqnarray}
U n_{i \uparrow} n_{i \downarrow} &\approx& U \bigl\{  n_{i \uparrow} \langle n_{i \downarrow} \rangle +
\langle n_{i \uparrow} \rangle n_{i \downarrow} - \langle n_{i \uparrow} \rangle \langle n_{i \downarrow} \rangle \bigr\} \nonumber \\
&=& - \sum_\sigma \frac{1}{2} U (M_i \sigma - N_i) n_{i \sigma}, 
\label{eq:U1}
\end{eqnarray}
where $M_i\equiv \langle n_{i \uparrow} - n_{i \downarrow} \rangle$ and $N_i \equiv \langle n_{i \uparrow} + n_{i \downarrow} \rangle$ are the spin antisymmetric and spin symmetric combinations of $\langle n_{i \sigma} \rangle$, respectively.
In the second line of Eq.~(\ref{eq:U1}), the constant term $U \langle n_{i \uparrow} \rangle \langle n_{i \downarrow} \rangle$ is neglected. 

We explore a self-consistent solution, in which the electron density is uniform while the spin polarization alternates between sublattices A and B, such that $N_A = N_B = N$ and $M_A = - M_B = M$. 
This is achieved by diagonalizing the $4\times4$ mean-field Hamiltonian 
\begin{equation} 
H_0 = \sum_{{\bf k}} \psi^\dagger_{{\bf k}} [{\bf h}({\bf k})\otimes \sigma_0 - h\tau_3\otimes \sigma_3] \psi^{\phantom\dagger}_{{\bf k}}, 
\label{eq:H0}
\end{equation}
and self-consistently computing the mean field order parameters $\langle n_{i \sigma}\rangle$. 
Here, $\psi^\dagger_{{\bf k}} = (c^\dagger_{A{\bf k}\uparrow}, c^\dagger_{B{\bf k}\uparrow}, c^\dagger_{A{\bf k}\downarrow}, c^\dagger_{B{\bf k}\downarrow})$, and and $\sigma_i$ are the Pauli matrices acting on the spin space.
$h = \frac{1}{2}UM$ is the staggered Zeeman field, and 
$\frac{1}{2}UN$ is absorbed into the chemical potential $\mu$. 
The self-consistency condition is given by
\begin{equation}
    \langle n_{\ell} \rangle = \frac{1}{L} \sum_{{\bf k}, \nu} 
    a_\nu^{\ell *}({\bf k}) a_\nu^\ell ({\bf k}) f (\xi_{\nu {\bf k}}-\mu), 
\end{equation}
where $\ell=(l,\sigma)$ labels sublattice, $l=A, B$, and spin, $\sigma = \uparrow,\downarrow$, 
$\xi_{\nu {\bf k}}$ are the $\bf k$-dependent energy eigenvalues of band $\nu$ with 
the corresponding eigenvectors given by $a^\ell_\nu({\bf k}) = \langle \mu{\bf k}|\ell{\bf k}\rangle$, and
$f(\xi_{\nu {\bf k}}-\mu)$ is the Fermi distribution function.

\begin{figure}[h] 
  \includegraphics[width=0.5\textwidth]{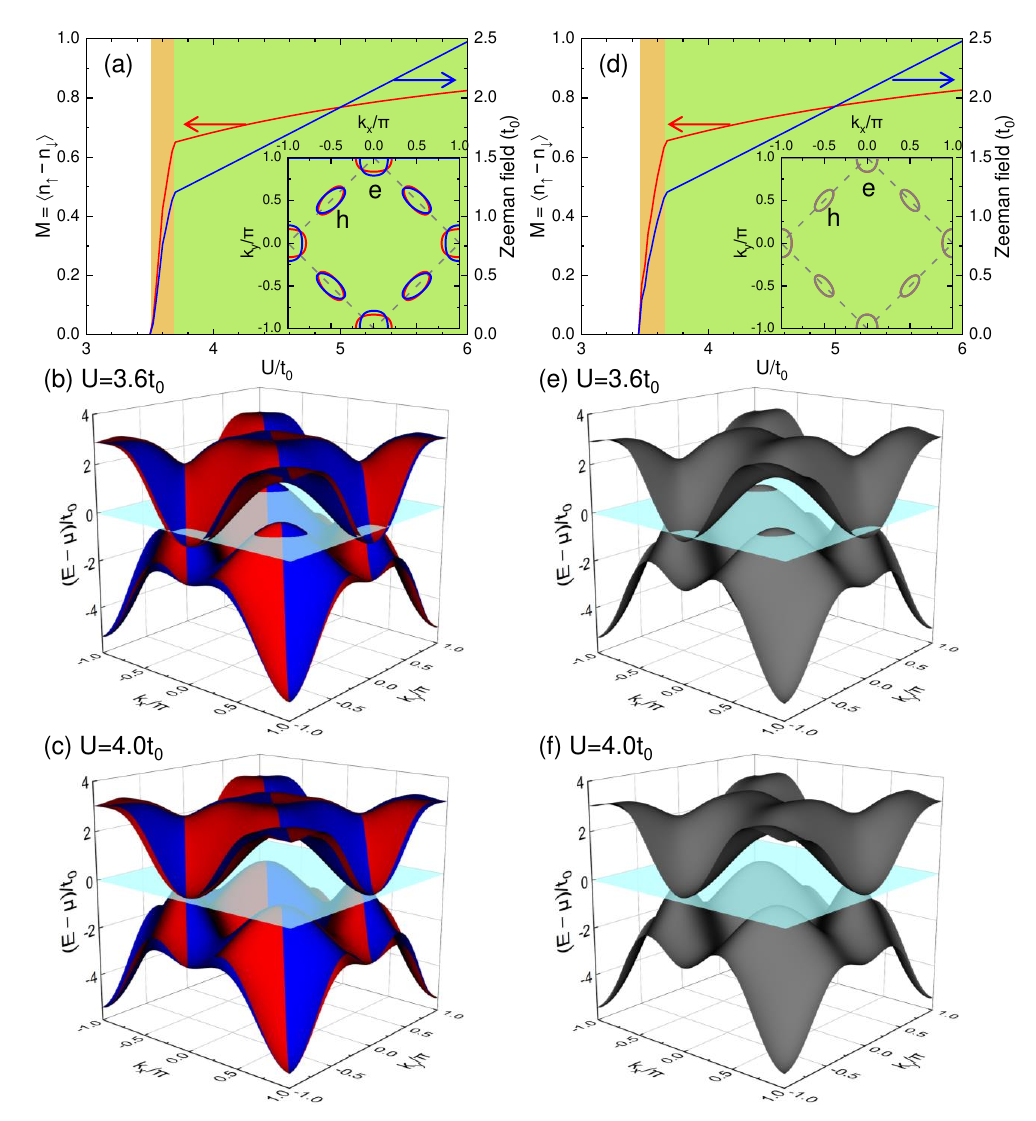} 
  \caption{Hartree-Fock results for altermagnetic (a-c) and antiferromagnetic (d-f) models. 
  (a, d) show the magnetization $M$ for the A-sublattice and the corresponding Zeeman field $h$. 
  (b, e) and (c, f) show the bandstructures in the metallic states with $U=3.6 t_0$ and the insulating states with $U=4.0t_0$, respectively. 
  Red and blue bands in (b, c) indicate the spin $\uparrow$ and $\downarrow$ components, respectively, showing $d_{x^2-y^2}$-type ${\bf k}$-dependent spin splitting, while gray bands in (e, f) indicate spin degeneracy. 
  The insets of (a, d) show the Fermi surfaces in the metallic states with $U=3.6t_0$, 
  with gray broken lines indicating the reduced Brillouin zone of the staggered magnetic state. The Fermi surfaces consist of hole pockets at $(\pm \pi/2, \pm \pi/2)$  and electron pockets at $(\pm \pi,0)$ and $(0,\pm \pi)$. 
  \label{fig:2}} 
\end{figure}

Fig.~\ref{fig:2} (a) shows the magnetization $M=\langle n_\uparrow-n_{\downarrow} \rangle$ for the A-sublattice (the magnetization on the B-sublattice is $-M$) as a function of the Coulomb $U$ repulsion. For the altermagnetic model, as shown in panel (a), we find a magnetic state with staggered spin polarization ($M$ on sublattice A, $-M$ on sublattice B) above a threshold of $U\gtrsim 3.5t_0$. Right above this critical $U$, we find a metallic state for a small range of $U$ as shown by the orange region, in which the sublattice magnetization $M$ and the corresponding Zeeman mean field $h$ rapidly rise with increasing $U$. For larger $U$, the bandstructure becomes gapped and the increase of $M$ and $h$ slows down significantly in the green shaded insulating region. The corresponding bandstructure in the altermagnetic state is illustrated in panel (b) for the metallic system at $U=3.6t_0$, and in panel (c) for an insulating system at $U=4t_0$. In both cases, the bandstructure has four different bands resulting from the sublattices and the spins, i.e. unlike in the antiferromagnetic state, the spin degeneracy is lifted, and the splitting between spin up (red) and down (blue) depends on the crystal momentum ${\bf k}$. As can be seen from the color shading in the figure, the splitting has a $d_{x^2-y^2}$ structure in momentum space. For the smaller $U$ metallic state in panel (b), the bands cross the Fermi level resulting in four electron pockets at the corners [$(\pi, 0)$, $(0,\pi)$, {\it etc}.] of the reduced Brillouin zone, and four hole pockets at the zone edges, as shown in the inset of Fig.~\ref{fig:2} (a). 
One sees that ${\cal C}_4$ rotational symmetry of the bandstructure and Fermi surface requires an additional spin-flip $\sigma\rightarrow\bar{\sigma}$ operation, due to the sign reversal of the $h\tau_3 \otimes \sigma_3$ term in Eq.~(\ref{eq:H0}) under the ${\cal C}_4$ operation.

In the ordered state, ${\cal PT}$ symmetry is broken, since for a general momentum ${\bf k}$, the degeneracy between ${\bf k}\uparrow$ and ${\bf k}\downarrow$ states is lifted. This can be unterstood from the form of the Hamiltonian in Eq.~(\ref{eq:H0}), since, under the ${\cal PT}$ operation, the $\frac{1}{2}[\epsilon_{AA}({\bf k})-\epsilon_{BB}({\bf k})]\tau_3$ term in ${\bf h}({\bf k})$ transforms into its negative value. For the altermagnetic model, this term is finite except along the diagonal $|k_x|=|k_y|$ and zone boundary $|k_x\pm k_y|=\pi$ due to the difference in $t_1$ and $t_2$ hoppings between the two sublattices.

The phase diagram shown in Fig.~\ref{fig:2} (d) for the corresponding translationally invariant model [Fig.~\ref{fig:1} (b)] with $\epsilon_{BB} = \epsilon_{AA}$ is similar, exhibiting an antiferromagnetic phase for $U\gtrsim 3.45t_0$ that evolves from a metallic state in the orange region to an insulating state in the green region. In contrast to the altermagnetic state, the corresponding bandstructure of this model shown in panels (e) and (f) is spin degenerate and lacks ${\cal C}_4$ symmetry since $t_1\neq t_2$. This can be seen from the Fermi surface for the $U=3.6t_0$ metallic state shown in the inset of panel (d). In this case, ${\cal PT}$ symmetry is preserved, because the $\frac{1}{2}[\epsilon_{AA}({\bf k})-\epsilon_{BB}({\bf k})]\tau_3$ term in ${\bf h}({\bf k})$ vanishes for the antiferromagnetic model.  

\section{Inelastic neutron scattering}



The dynamic spin susceptibility that determines the neutron scattering intensity is given by
\begin{eqnarray}
    \chi_{ij}({\bf q}, i\omega_m) &=& \sum_{l, l'}\int^{1/T}_0 d\tau e^{i\omega_m\tau}\langle {\cal T}_\tau S_i^{l}({\bf q},\tau)S_j^{l'}({\bf -q}, 0)\rangle\,,\nonumber
\end{eqnarray}
for transferred momentum ${\bf q}$ and bosonic Matsubara frequencies $\omega_m=2m\pi T$ with temperature $T$. We will consider both  scattering with positive chirality given by $\chi_{+-}$ and scattering with negative chirality given by $\chi_{-+}$. These susceptibilities are calculated with the spin-flip operators $S_+^{l}({\bf q}) = \sum_{\bf k} \psi^\dagger_{l\uparrow}({\bf k+q})\psi^{\phantom\dagger}_{l\downarrow}({\bf k})$ and $S_-^{l}({\bf q}) = \sum_{\bf k} \psi^\dagger_{l\downarrow}({\bf k+q})\psi^{\phantom\dagger}_{l\uparrow}({\bf k})$. In the multi-orbital RPA approximation \cite{Kubo2007, Graser2009}, $\chi_{ij}$ is calculated from an orbital dependent spin-susceptibility tensor
\begin{eqnarray}
    \chi^{\ell_1\ell_2}_{\ell_3\ell_4}({\bf q}, i\omega_m) = \left\{\chi_0({\bf q}, i\omega_m)[1-{\cal U}\chi_0({\bf q}, i\omega_m)]^{-1} \right\}^{\ell_1\ell_2}_{\ell_3\ell_4}\,.\nonumber
\end{eqnarray}
Here, ${\cal U}$ is a 16$\times$16 matrix containing the on-site Coulomb interaction $U$ with matrix elements ${\cal U}^{\ell_1\ell_2}_{\ell_3\ell_4} = U\delta_{l_1l_2}\delta_{l_1l_3}\delta_{l_1l_4} [\delta_{\sigma_1\bar{\sigma_2}}\delta_{\sigma_1\sigma_3}\delta_{\sigma_1\bar{\sigma_4}} - \delta_{\sigma_1\sigma_2}\delta_{\sigma_1\bar{\sigma_3}}\delta_{\sigma_1\bar{\sigma_4}}]$, and the matrix elements of the bare susceptibility $\chi_0$ are given by
\begin{eqnarray}
    (\chi_0)^{\ell_1\ell_2}_{\ell_3\ell_4}({\bf q}, i\omega_m) = &-&\frac{T}{L}\sum_{{\bf k}, \varepsilon_n} G_0^{\ell_3\ell_1}({\bf k+q},i\varepsilon_n+i\omega_m) \nonumber\\ 
    &\times&G_0^{\ell_2\ell_4}({\bf k}, i\varepsilon_n) \,,
\end{eqnarray}
with $G_0^{\ell\ell'}({\bf k}, i\varepsilon_n)$ the bare Green's function
\begin{eqnarray}
    G_0^{\ell\ell'}({\bf k}, i\varepsilon_n) &=& \sum_\nu \frac{a^\ell_\nu({\bf k}) a^{\ell'*}_\nu({\bf k})}{i\varepsilon_n-\xi_{\nu{\bf k}} + \mu}
\end{eqnarray}
and $\varepsilon_n=(2n+1)\pi T$ the fermionic Matsubara frequencies. 
Carrying out the usual analytic continuation of Matsubara frequencies $\omega_m$ to the real frequency $\omega$ axis, the physical spin susceptibility $\chi_{ij}({\bf q}, \omega)$ is then obtained as
\begin{eqnarray}
    \chi_{ij}({\bf q}, \omega) = &&\sum_{l,l'}\sum_{\alpha,\beta,\gamma,\delta} \sigma^i_{\alpha\beta} \sigma^j_{\gamma\delta} \\
    &\times&\chi^{l\alpha,l\beta}_{l'\gamma,l'\delta}({\bf q}, i\omega_m\rightarrow \omega+i\eta)\,.\nonumber
\end{eqnarray}
Here, the operators $\sigma^+ = \frac{1}{2}(\sigma_1+i\sigma_2)$ and $\sigma^- = \frac{1}{2}(\sigma_1-i\sigma_2)$ with the Pauli matrices $\sigma_i$, and $\eta$ is a positive infinitesimal. 

\begin{figure}[ht!] 
\vspace{0.25cm}
  \includegraphics[width=0.5\textwidth]{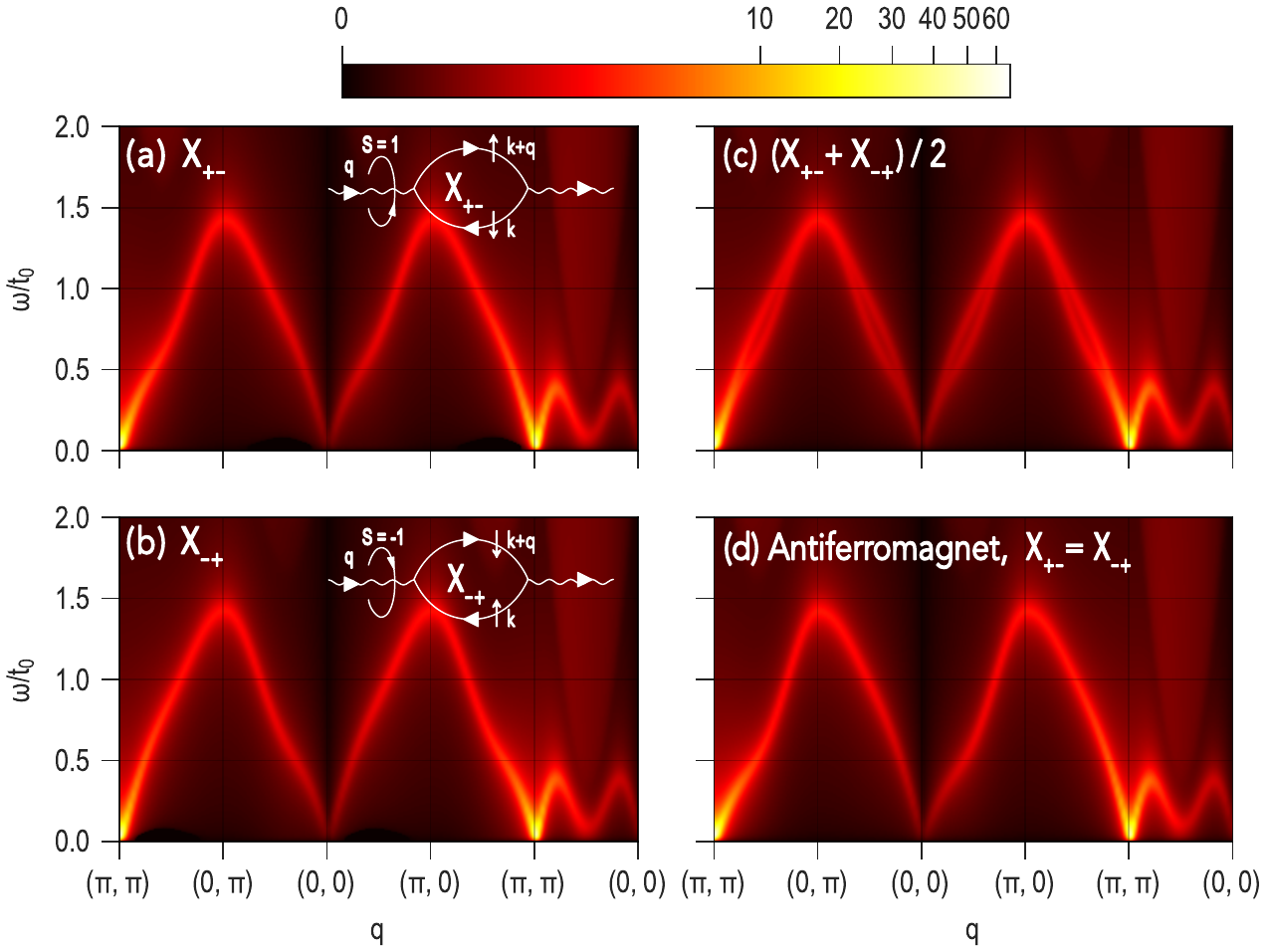} 
  \caption{RPA spin susceptibility for the insulating state at $U=4t_0$: (a) $\chi''_{+-}({\bf q}, \omega)$ and (b) $\chi''_{-+}({\bf q}, \omega)$, and (c) their average, for the altermagnetic model shown in Fig.~\ref{fig:1}(a), and (d) $\chi''_{+-}({\bf q}, \omega)=\chi''_{-+}({\bf q}, \omega)$ for the orthorhombic antiferromagnetic model shown in Fig.~\ref{fig:1}(b).  \label{fig:3}}
\end{figure}

In Fig.~\ref{fig:3} we show the imaginary part of the dynamic spin susceptibilities $\chi''_{+-}({\bf q}, \omega)$ [panel(a)] and $\chi''_{-+}({\bf q}, \omega)$ [panel (b)] and their average [panel (c)] for the altermagnetic state calculated in the insulating region for $U=4t_0$ along high-symmetry directions in the Brillouin zone. As a comparison, panel (d) shows the spectrum obtained for the antiferromagnet with orthorhombic anisotropy. For all cases, the spin-wave dispersion has minima at ${\bf q}=0$, $(\pi, \pi)$ and $(\pi/2,\pi/2)$ with the largest intensity at $(\pi, \pi)$ because of the staggered magnetic ordering. A closer inspection of the detailed dispersion near ${\bf q}=0$, however, reveals important differences: Due to the orthorhombic anisotropy in the hopping amplitudes, the dispersion along $q_x$ from ${\bf q}=0$ to $(\pi,0)$ is different from the dispersion along $q_y$ from ${\bf q}=0$ to $(0, \pi)$. For the altermagnetic model, this difference is reversed between the neutron scattering response with positive chirality described by $\chi_{+-}$ shown in panel (a) and the response with negative chirality shown by $\chi_{-+}$ in panel (b). This difference is similar to that observed previously in calculations for an effective Heisenberg model of RuO$_2$ in Ref.~\cite{Smejkal2022c}. As a result of the chirality dependent spin-wave dispersion, the average of $\chi_{+-}$ and $\chi_{-+}$ shown in panel (c) displays two branches dispersing out of ${\bf q}=0$. For the antiferromagnetic case shown in panel (d) obtained for the model with orthorhombic anisotropy, a similar asymmetry between $q_x$ and $q_y$ is seen. For this case, however, the spectrum does not depend on chirality, i.e. $\chi_{+-} = \chi_{-+}$, and the total scattering intensity shows only a single branch. 

\begin{figure}[ht!] 
\vspace{0.25cm}
  \includegraphics[width=0.5\textwidth]{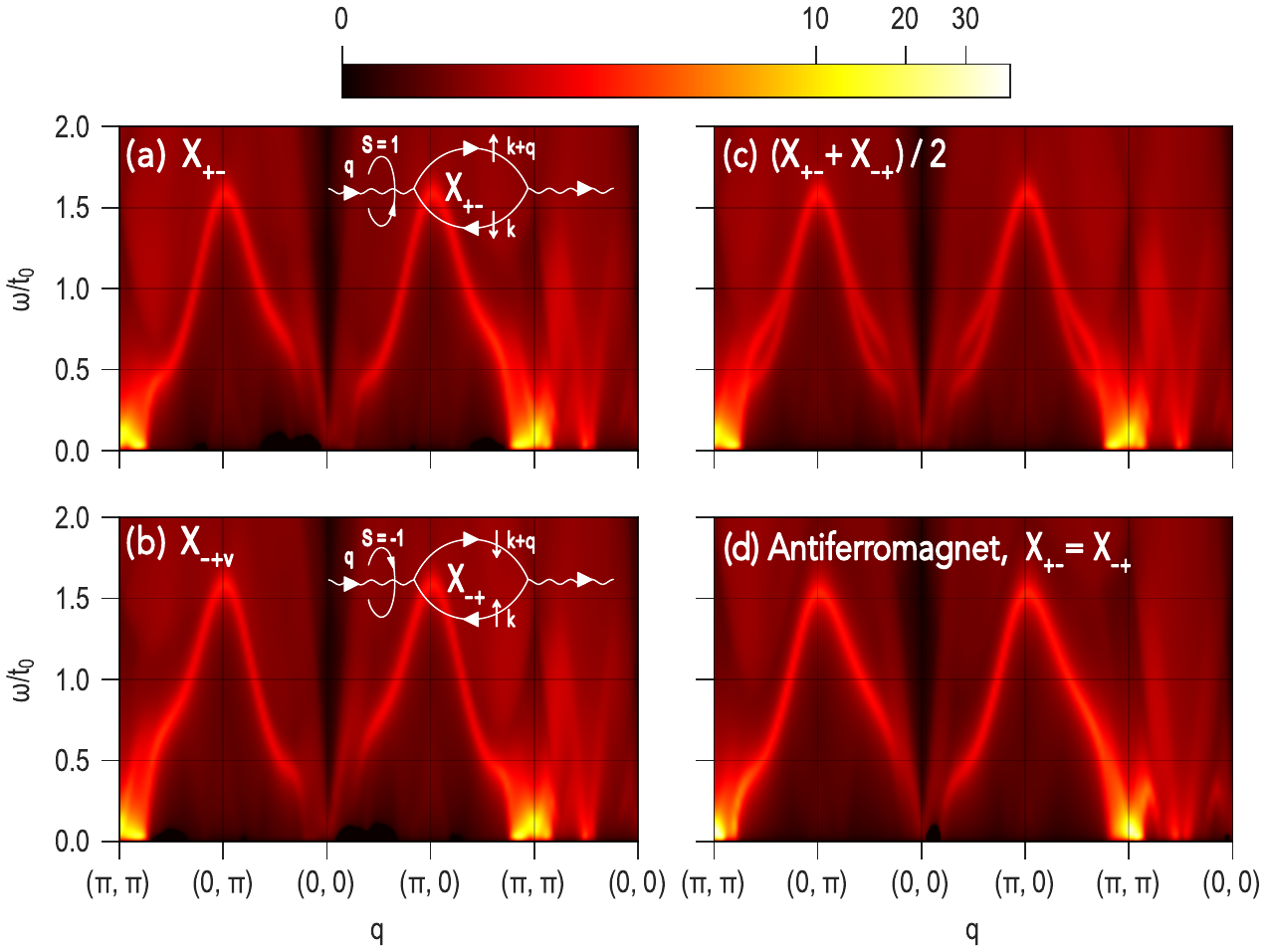} 
  \caption{RPA spin susceptibility for the metallic state at $U=3.6t_0$: (a) $\chi''_{+-}({\bf q}, \omega)$ and (b) $\chi''_{-+}({\bf q}, \omega)$, and (c) their average, for the altermagnetic model shown in Fig.~\ref{fig:1}(a), and (d) $\chi''_{+-}({\bf q}, \omega)=\chi''_{-+}({\bf q}, \omega)$ for the orthorhombic antiferromagnetic model shown in Fig.~\ref{fig:1}(b). \label{fig:4}
  } 
\end{figure}

Fig.~\ref{fig:4} shows the corresponding spectrum obtained for the metallic state at $U=3.6t_0$. In this case, the overall intensity drops to about half of that for the insulating state and the increased Landau damping due to the metallic character is obvious. Similar to the insulating case (Fig.~\ref{fig:3}), however, a clear asymmetry in the low energy spin excitations is visible near ${\bf q}=0$. For the altermagnetic case [panels (a) and (b)], this asymmetry is again reversed between different chiralities, and the average of $\chi_{+-}$ and $\chi_{-+}$ has two distinct branches for energies up to $\sim t_0$ [panel (c)]. 

In the RPA approach, the chirality dependence of the response can be traced to an asymmetry in the propagation between spin $\uparrow$ and spin $\downarrow$ particles. As shown in the insets of Figs.~\ref{fig:3}(a) and \ref{fig:4}(a), the neutron scattering process with positive chirality involves a spin $S=1$ particle-hole excitation where momentum ${\bf q}$ is transferred to a spin $\uparrow$ electron. Conversely, as shown in the insets of Figs.~\ref{fig:3}(b) and \ref{fig:4}(b), a scattering process with negative chirality involves a spin $S=-1$ particle-hole excitation where momentum ${\bf q}$ is transferred to a spin $\downarrow$ electron. Since spin $\uparrow$ and spin $\downarrow$ electrons propagate mainly on different sublattices, the difference in the magnetic excitations described by $\chi_{+-}$ and $\chi_{-+}$ can be traced to the differences in the electronic structure between the two sublattices. 

\section*{Conclusion}
We have introduced a microscopic two-sublattice Hubbard model with alternating, ${\cal C}_4$ symmetry breaking orthorhombic anisotropy in the hopping parameters, which provides a minimal framework to study altermagnetism. Using an unrestricted Hartree-Fock approximation, we have found an altermagnetic state with staggered magnetization and spin-split bandstructure, which evolves from a metallic state at small $U$ to an insulating state at larger $U$. We have used this model and an RPA approximation of the dynamic spin susceptibility to study the inelastic neutron scattering response in the altermagnetic state, and compared this response with that of a corresponding conventional antiferomagnetic model with orthorhombic anisotropy, for both the metallic and insulating states.  As one would expect, we have found that the magnetic spectrum described by $\chi''({\bf q}, \omega)$ breaks ${\cal C}_4$ symmetry for both these models, and in the case of the altermagnetic model, depends also on its chirality, described by either $\chi''_{+-}$ or $\chi''_{-+}$, for ${\bf q}$ along certain directions in the Brillouin zone. For the conventional antiferromagnetic state, the magnetic excitation spectrum is chirality-degenerate. The observation of chirality-split magnetic excitations in inelastic neutron scattering experiments may therefore be used as evidence for altermagnetism.

\begin{acknowledgements}
We acknowledge useful discussions with Andrew Christianson, Hu Miao, and Pyeongjae Park. This work was supported by the U.S. Department of Energy, Office of Science, Basic Energy Sciences, Materials Sciences and Engineering Division. This manuscript has been authored by UT-Battelle, LLC, under Contract No. DE-AC0500OR22725 with the U.S. Department of Energy.  The United States Government retains and the publisher, by accepting the article for publication, acknowledges that the United States Government retains a nonexclusive, paid-up, irrevocable, world-wide license to publish or reproduce the published form of this manuscript, or allow others to do so, for the United States Government purposes. The Department of Energy will provide public access to these results of federally sponsored research in accordance with the DOE Public Access Plan (\url{http://energy.gov/downloads/doe-public-access-plan}).
\end{acknowledgements}

\bibliographystyle{revtex4}
%

\end{document}